\documentclass[useAMS,usenatbib,usegraphicx]{mn2e}

\usepackage{multirow}
\usepackage{url}
\usepackage{threeparttable}
\usepackage{graphicx}
\usepackage[T1]{fontenc}
\usepackage{aecompl}
\usepackage{color}
\usepackage{ulem}

\usepackage{breqn}

\title[Long time-scale variability of LMXB]{Long time-scale variability of X-ray binaries with late type giant companions}

\author[Filippova et al.]{E. Filippova$^{1,2}$\thanks{E-mail: kate@iki.rssi.ru}, M. Revnivtsev$^2$, E. R. Parkin$^{3}$\\
$^1$ ISDC, University of Geneva, Switzerland\\
$^2$ Space Research Institute, Moscow, Russia\\
$^3$ Research School of Astronomy and Astrophysics, Australian National University, Canberra, ACT 2611, Australia
}

\date{Accepted.}

\pagerange{\pageref{firstpage}--\pageref{lastpage}} \pubyear{2013}

\def\LaTeX{L\kern-.36em\raise.3ex\hbox{a}\kern-.15em
    T\kern-.1667em\lower.7ex\hbox{E}\kern-.125emX}

\begin{document}

\maketitle

\label{firstpage}

\begin{abstract}
In this paper we propose and examine a physical mechanism which can lead to the generation of noise in the mass accretion rate of low mass X-ray binaries on time-scales comparable to the orbital period of the system. We consider modulations of mass captured by the compact object from the companion star's stellar wind in binaries with late type giants, systems  which usually have long orbital periods. We show that  a hydrodynamical interaction of the wind matter within a binary system even without eccentricity results in variability of the mass accretion rate with characteristic time-scales close to the orbital period. The cause of the variability is an undeveloped turbulent motion (perturbed motion without  significant vorticity) of wind matter near the compact object. Our conclusions are supported by 3D simulations with two different hydrodynamic codes based on Lagrangian and Eulerian approaches -- the SPH code GADGET and the Eulerian code PLUTO. 
In this work we assume that the wind mass loss rate of the secondary is at the level of $(0.5-1)\times10^{-7}~M_\odot$/year, required to produce observable variations of the mass accretion rate on the primary. This value is higher than that, estimated for single giant stars of this type, but examples of even higher mass loss rate of late type giants in binaries do exist.
	Our simulations show that the stellar wind matter intercepted by the compact object might create observational appearances similar to that of an accretion disc corona/wind and could be detected via high energy resolution observations of X-ray absorption lines, in particular,  highly ionized ions of heavy elements.
\end{abstract}

\begin{keywords}
X-rays: binaries -- accretion: stellar wind -- stars: mass-loss -- methods: numerical 
\end{keywords}

%\tableofcontents

\section{Introduction}

Low mass X-ray binaries (LMXBs) exhibit variability in their fluxes across a wide range of time-scales. It is generally accepted that in these binaries the matter is accreted by a compact star via an accretion disc which is the dominant source of emission in the energy bands from optical to X-rays \citep*{shakura_1973, frank_2002}. 

The currently favoured model for the X-ray flux variability of LMXBs is that of propagating fluctuations \citep*{lyubarskii_1997,churazov_2001,uttley_2001,kotov_2001,ingram_2013}. In this model a fluctuating viscosity generates variability in the mass accretion rate at different radii of the accretion flow. The variations in the mass accretion rate propagate into the innermost regions of the accretion disc and appear as variability in the X-ray (and optical/IR) luminosity of the binary. The self-similar nature of such variations leads to the generation of flux variability with a power law type power density spectrum (PDS)\cite[see e.g.][]{lyubarskii_1997}.
The smallest Fourier frequency at which the viscosity fluctuations in the disc can generate a power law type of flux variability, $f_{\rm visc}$, corresponds to the viscous time-scale at the outer radius of the accretion disc.  Theoretical estimates for typical parameters of the accretion disc in LMXBs show that it should be around $f_{\rm visc}\sim (0.005\--0.07 )f_{\rm orb}$, where $f_{\rm orb}$ is an orbital frequency of the system. At smaller frequencies the accretion disc can generate only random noise in the mass accretion rate, which should result in a flattening of the flux variability power spectrum.

An observational test of this model was presented by \cite{gilfanov_2005}. These authors analysed the X-ray flux variability for a set of LMXBs and demonstrated that their power density spectra indeed have a power law shape across a wide range of time-scales. It was found that binary systems with orbital periods greater than $P_{\rm orb}\sim 18\, {\rm h}$ clearly demonstrate a flattening of their PDSs at Fourier frequencies below $\sim 0.1\times f_{\rm orb}$, while LMXBs with smaller orbital periods ($P_{\rm orb}<4\,{\rm h}$ in this work) have PDSs in which the power law extends to much lower frequencies, and some (mostly ``dippers'', i.e. binaries, which due to high inclination demonstrate ``dips'' in their fluxes related to partial obscuration of the emitting regions at some orbital phases) have breaks around $\sim f_{\rm orb}$.
 
It is clearly seen that systems in the work of \cite{gilfanov_2005}  are separated into two groups based on the type of the secondary companion: with giant/sub-giant stars (orbital periods $P_{\rm orb}>18$ h) and with main sequence/white dwarf stars. This suggests that the presence of a stellar wind from a giant companion might have an influence on the resulting mass accretion rate variability in binaries.  The aim of the present work is to explore the temporal behaviour of accretion from the stellar wind of a late type giant in a long period LMXB and estimate the observational appearance of the stellar wind in such systems. 
 Numerical simulations are a suitable tool to investigate the influence of the stellar wind accretion on the compact object's mass accretion rate variability.  
Although a lot of papers are devoted to simulations of wind accretion in X-ray binaries \citep*{biermann_1971,matsuda_1987,theuns_1996,nagae_2004,valborro_2009}  variability of the mass accretion rate has not been analysed in details.

The structure of this paper is as follows: in section \ref{physics} we describe our model  of accretion through the stellar wind of late type giants and its physical basis. In section \ref{numsim}  we describe numerical simulations of the model. In section \ref{results} we discuss obtained results and show that the existence of accretion from a stellar wind, even in the case of zero eccentricity for the binary motion, can lead to a flattening of the PDS of long period LMXBs at low Fourier frequencies. 
In section \ref{stellar_wind} we show a possible observational appearance of the stellar wind in such systems and close with conclusions in section \ref{conclusions}.

\section{Model description}\label{physics}

\begin{figure}
\centerline{
\includegraphics[width=\columnwidth]{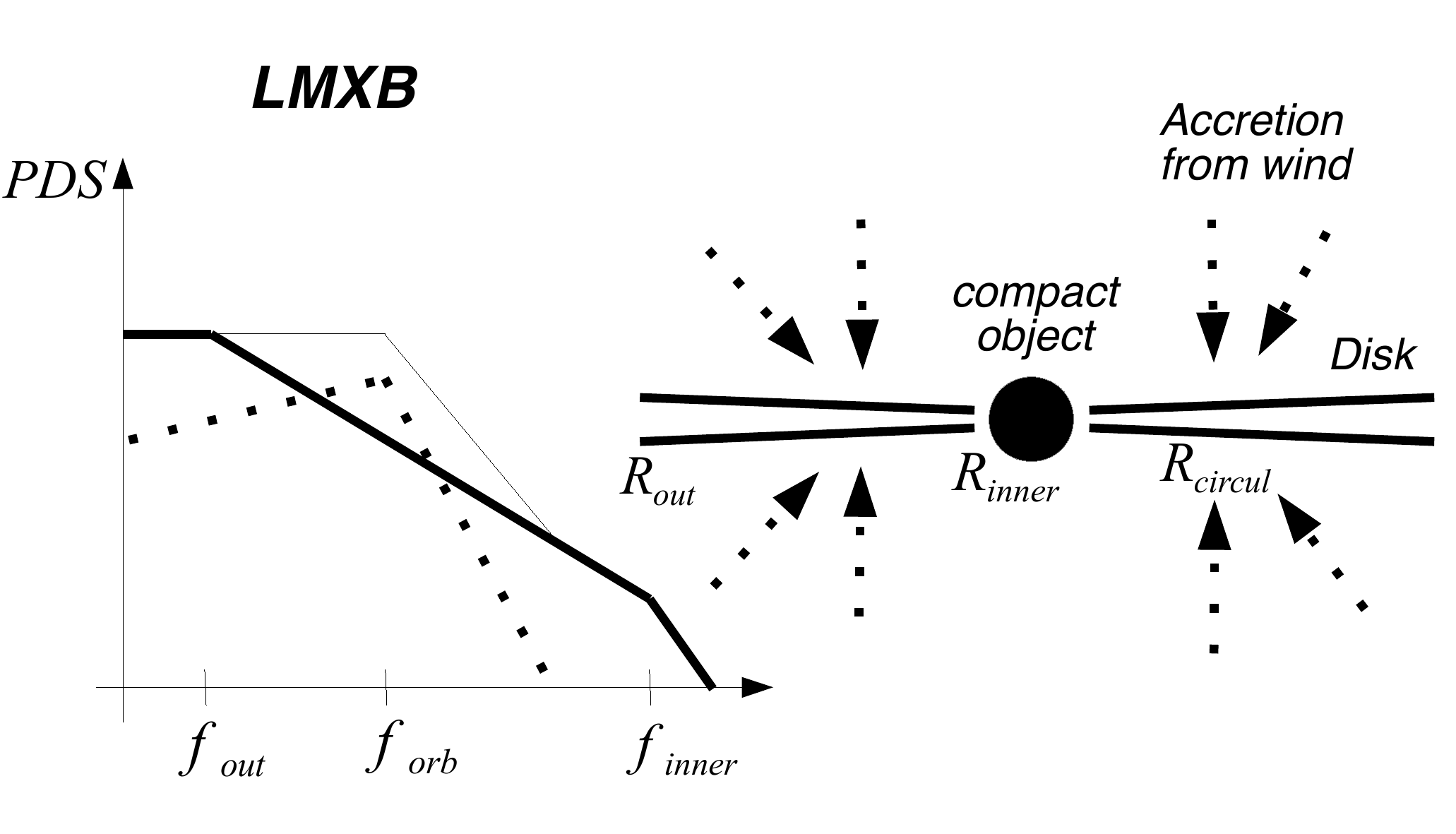}
 }
\caption{A schematic view of formation of the variability of LMXB's X-ray luminosity  as a sum of mass accretion rate variability from a disc (thick line) and from the stellar wind of a giant companion (dotted line).  The radii in the right panel have the following definitions: $R_{\rm out}$ and $R_{\rm in}$ are the outer and inner disc radii, respectively, and $R_{circul}$ is the circularization radius of the accreted wind matter. The resulted spectrum is shown by thin line on the plot.
}\label{accretion_from_wind}

\end{figure}

A schematic view of accretion in an LMXB with a late type giant companion is depicted in Fig. \ref{accretion_from_wind}.
According to this idea we will show that, in spite of the assumed absence of eccentricity in the binary system, the accretion rate variability  from the stellar wind has some excess of power at time-scales comparable to the orbital period of the system. Variability at these relatively short  time-scales (in comparison to the viscous time-scale of the accretion disc at  the wind circularization radius) should be transported to the innermost parts of the accretion flow, presumably via coronal flow above the accretion disc.  Observational evidence of this flow has been presented by a lot of authors, e.g. \cite*{cottam_2001, jimenez_2003, kallman_2003,boirin_2004,boirin_2005,bayless_2010}.

\subsection{Parameters of the binary}

We perform our calculations for the case of the brightest low mass X-ray binary, Sco X-1. Parameters of this system are listed in Table 1.

\begin{table}
\begin{minipage}{\columnwidth}
\caption{Adopted parameters for ScoX-1}
\begin{tabular}{l|l}
\hline
Mass ratio, $M_{\rm 2}/M_{\rm 1}$ &0.3 \citep{steeghs_2002}\\
&\\
Compact object &\\
mass, $M_{\rm 1}$ &$ 1.4 M_{\odot}$\\
&\\
Distance &\\
between stars&$3.05\times10^{11}$ cm \\
&\\
Eccentricity & 0 \\
&\\
Orbital period & 0.788 d = 18.92 h \\
&\\
Optical star type & earlier then G5, sub-giant\\
& \citep{bandy_1997,bandy_1999}\\
&\\
Inclination angle & $40^{\circ}$ \\
&\citep*{fomalont_2001}\\
&\\
Radius of optical star& $R_{\rm 2,L}=8.5\times10^{10}$ cm\\
&\\
"Accretion radius"&\\ 
of compact star& $6.28\times10^{10} {\rm cm} \sim 0.42 R_{\rm 1,L}$\\
&\\
Stellar wind velocity&$\mathrm{380\, km\, s^{-1}}$\\
&\\
Mass loss rate &$\mathrm{5\times 10^{-8}, 10^{-7} M_{\odot}}$yr$^{-1}$\\
of the optical star&\\

\hline
\end{tabular}
\end{minipage}
\end{table}\label{parameters_scox1}

From the parameters in Table 1 we determine the position of the compact and optical stars as $7\times 10^{10}$ cm and $23.5\times10^{10}$ cm from the centre of mass, respectively. We assume that  optical star fills its Roche lobe, with the Roche lobe radius calculated in the usual way.
Furthermore, we assume that the velocity of the stellar wind corresponds to the star's escape velocity, $v_{\rm wind}=\sqrt{2\times GM_{\rm 2}/R_{\rm 2,L}}$.
For the Roche lobe radius of the optical star $R_{\rm 2,L}=8.5\times10^{10}$ cm we have $v_{\rm wind}\sim380\, {\rm km\,s^{-1}}$. 
This velocity may appear to be too large for a late type giant which is typically several ten's ${\rm km\,s^{-1}}$. However, one should bear in mind that the optical companion in Sco X-1 is smaller than a typical late type giant star due to the removal of its outer layers via accretion on to the compact object. In addition, observations of  X-ray binaries with mass losing giants sometimes show high stellar wind velocities similar to the value we estimate above \citep{chakrabarty_1998}.

In our simulations the initial temperature of the stellar wind is 11600 K which then quickly cools down to $\sim 7000$ K due to adiabatic expansion at the radius of the Roche lobe (which plays the role of the optical star's surface).

\subsection{Mass loss rate of optical star} \label{mass_loss_rate}

 The mass loss rate from late type giants in compact binaries is not accurately known. Simple scaling laws for single stars predict a relatively low level of mass loss rate for giants of G-M type \citep{reimers_1975,reimers_1977,schroder_2005} but observations of late type giants in binaries indicate values up to $\mathrm{\sim 10^{-6} M_{\odot}\,yr^{-1}}$
(e.g. GX 1+4 - \cite*{chakrabarty_1998}, and, CH Cyg - \cite{faraggiana_1969,skopal_2002,bogdanov_2008}).
 One reason for this enhancement in the wind mass loss rate can be the powerful X-ray illumination from the secondary star \citep*{arons_1973,basko_1973,iben_1997}. Alternatively, the field of forces, which are generating the stellar wind in the system, can significantly alter the geometry of the Roche lobe compared to the standard picture \citep{dermine_2009}.

Our approach, therefore, is to estimate a mass loss rate based on the value that would be required to significantly influence the mass accretion rate onto the compact object. 
If we want the variability of the mass accretion rate from the stellar wind to be detectable in the observed light curve then the mass accretion rate from the wind $\dot{M}_{\rm wind,acc}$ should be at least comparable to the total observed mass accretion rate, $\dot{M}_{\rm acc}$. 
The mass loss rate of the star is  $\dot{M}_{\rm wind}=1/\beta \dot{M}_{\rm wind,acc}$, where $\beta$ is an efficiency of wind capture by the compact star. Our simulations show that for our setup $\beta \sim 30-40$ per cent. 

The total mass accretion rate of the compact star, $\dot{M}_{\rm acc}$, can be estimated from the X-ray luminosity of the system, $L_{\rm x}=\eta \dot{M}_{\rm acc} c^2$, where $\eta\sim 0.1\--0.2$ is the accretion efficiency \citep{sibgatullin_2000}. For the X-ray luminosity of Sco X-1 $L_{\rm x} \sim2.7\times 10^{38} {\rm erg\, s^{-1}}$ \citep{gilfanov_2005} we have $\dot{M}_{\rm acc}\sim (2\--5) \times 10^{-8} \mathrm{M_{\odot}\,yr^{-1}}$. 
 
Therefore, in order to make a significant contribution to the total mass accretion rate, the wind mass loss rate of the optical star should be $\dot{M}_{\rm wind} \sim 0.5 /( \beta =0.3) \times \dot{M}_{\rm acc} \sim (3\--8)\times 10^{-8} \mathrm{M_{\odot}\,yr^{-1}}$.
For our subsequent simulations we assume  $\dot{M}_{\rm wind}=5\times 10^{-8}$ and  $10^{-7} \mathrm{M_{\odot}\,yr^{-1}}$.

\subsection{Radiative cooling}\label{rad_cooling}

In our simulations we do not take into account radiative cooling of the wind matter heated by shocks around the compact object. Our assumption is that cooling is compensated for by X-ray emission from the compact object via Compton heating. 
To justify this point we estimate the hot plasma cooling function using the APEC model in XSPEC.

For Compton heating we use the formula:

\[
{d E_{\rm Compt}\over{dV dt}}=4 \pi n_e {{\sigma_t}\over{m_e c^2}} \int d\nu J_{\nu} (h_p\nu -4kT_e)= 
\]
\[
 ={n_e \sigma_t L \over{4\pi R^2 m_ec^2}}\left(<h\nu>-4kT_e\right) ~ {\rm erg\,s^{-1}\,cm^{-3}},
\]

\noindent where $\sigma_t=6.65\times10^{-25}$ cm$^{2}$ is the Thompson scattering cross-section, 
$L$ is the central source luminosity, $R$ is the distance from the source, and $<h\nu>$ is the mean photon energy.

We assume that the plasma is illuminated by black body emission with a temperature of $kT_{\rm bb}=2.4$ keV \citep{dai_2007}, $L\sim 2.7\times10^{38}{\rm erg\,s^{-1}}$, and $<h\nu>=2.7kT_{\rm bb}=6.48\, {\rm keV}$. 

For typical densities and temperatures in our case, $n=10^{12}$ cm$^{-3}$ and  $kT_e\sim 0.4\,{\rm keV}$ (Fig. \ref{profili}, stellar wind mass loss rate is $\dot{M}=10^{-7}  \mathrm{M_{\odot}\,yr^{-1}}$). Assuming solar abundance,  the cooling rate for the plasma, $\Lambda =2\times 10^{-23} {\rm erg\,cm^3\,s^{-1}}$ \citep*{filippova_2008}. Combining these values,  we have $dE_{\rm line~ em}/(dV dt)\sim \Lambda\times n^2=20\, {\rm erg\,cm^{-3}\,s^{-1}}$  which is comparable with the heating rate $dE_{\rm Compt}/(dV dt)\sim 13.5\, {\rm erg\,cm^{-3}\,s^{-1}}$.  
Thus  it is reasonable to neglect radiative cooling in our simplest approximation.

\section{Numerical simulations}\label{numsim}

\begin{figure*}

\includegraphics[width=\columnwidth]{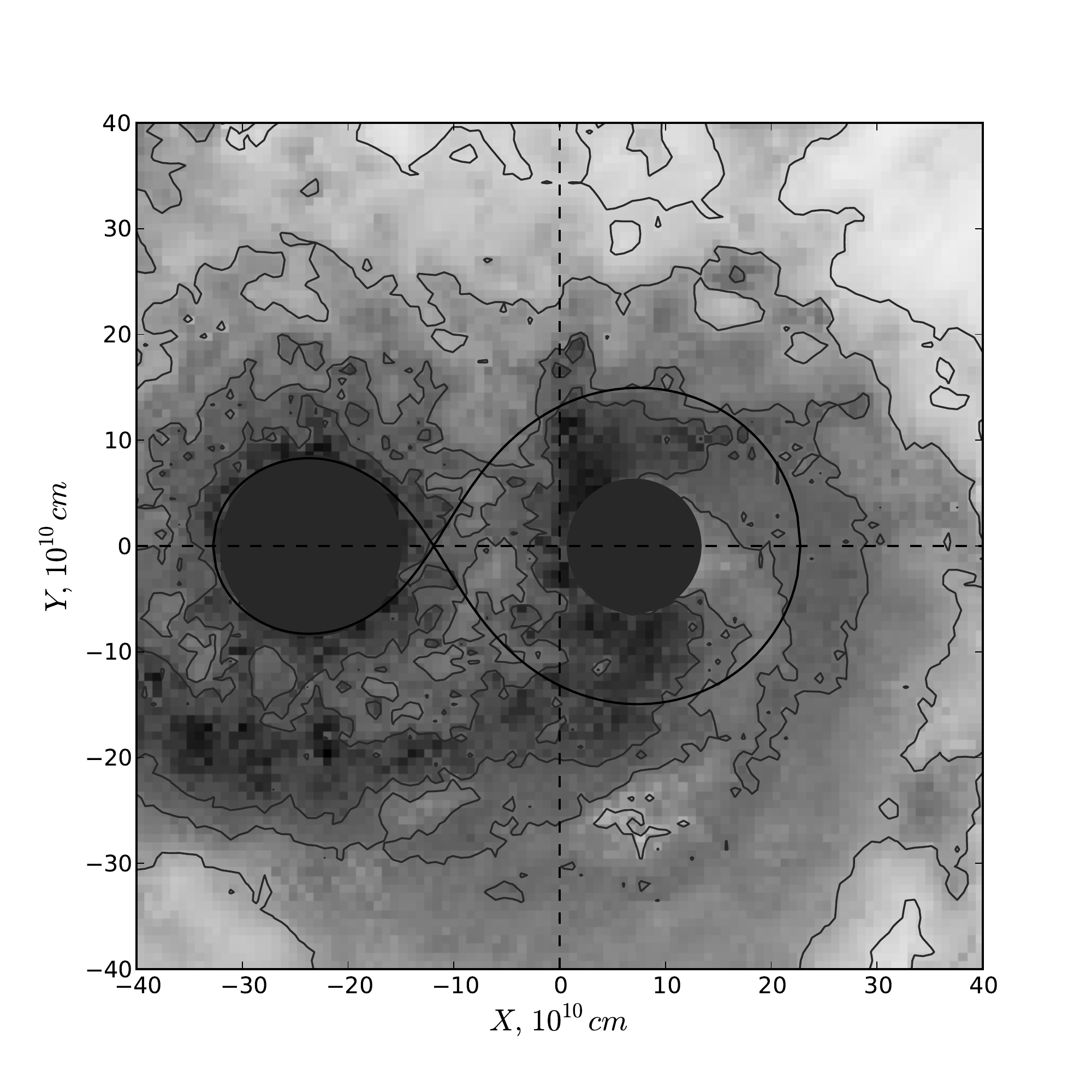}
\includegraphics[width=\columnwidth]{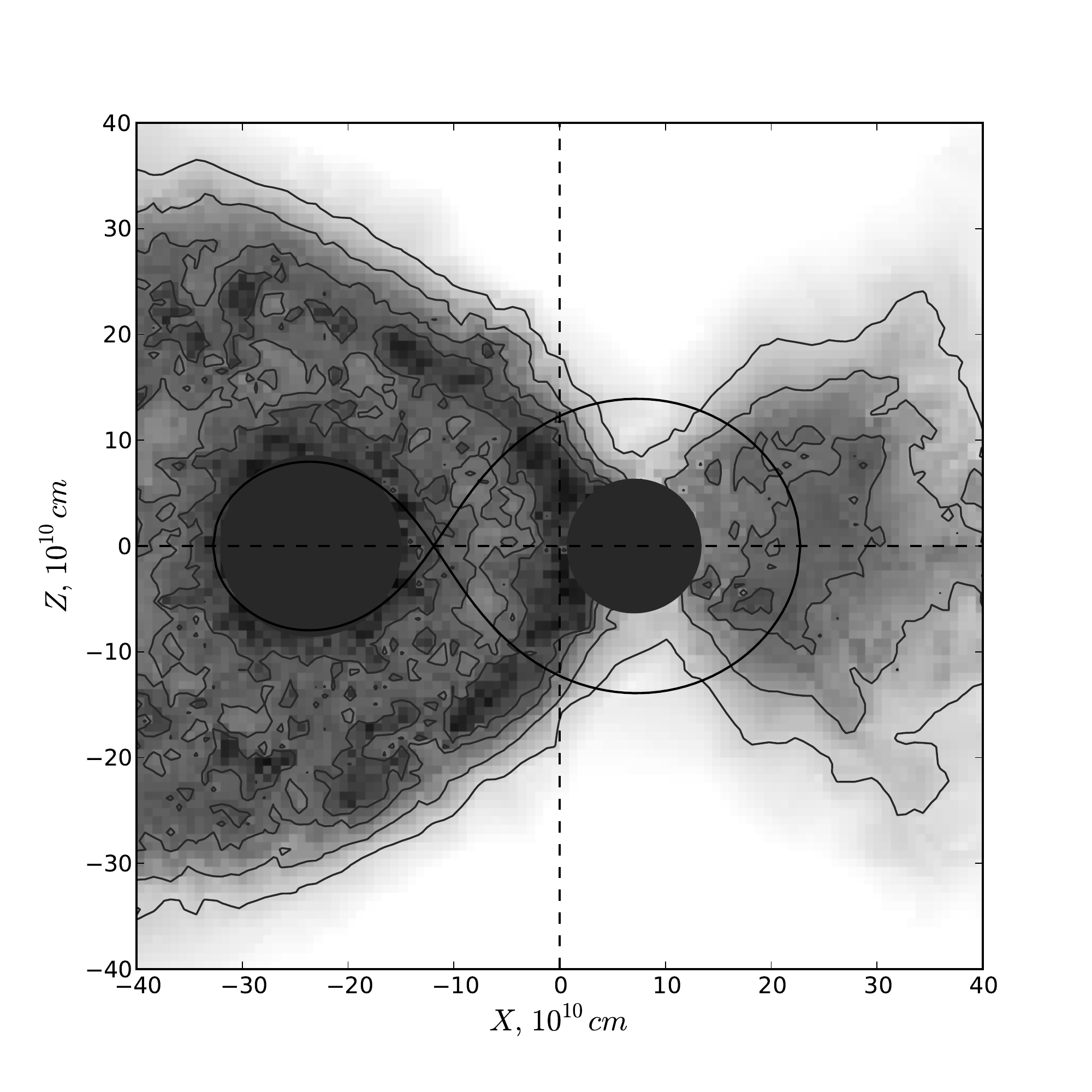}

\includegraphics[trim=1.8cm 1cm -0.7cm 0cm, clip=true,width=\columnwidth]{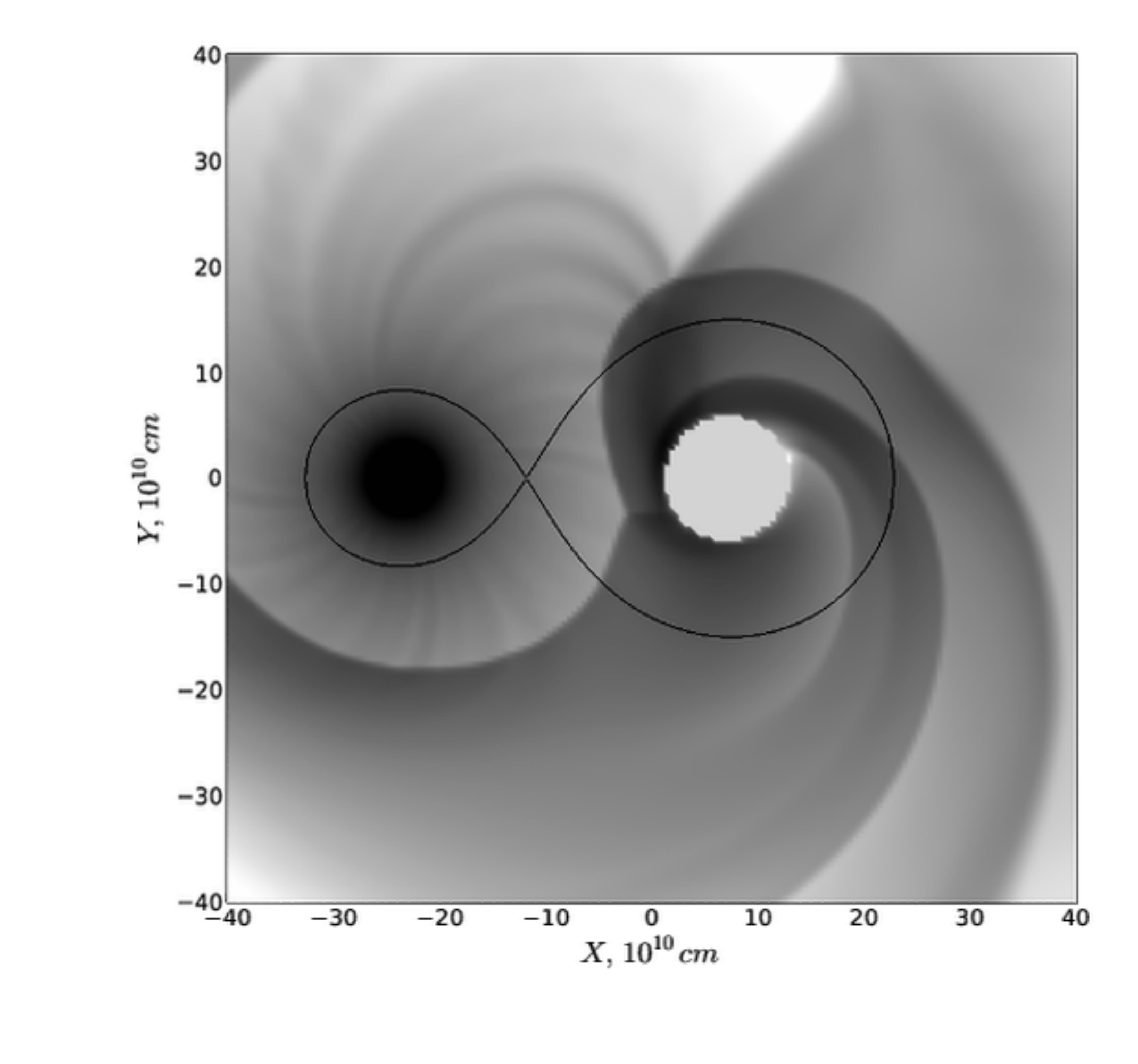}
\includegraphics[trim=1.8cm 1cm -0.7cm 0cm, clip=true, width=\columnwidth]{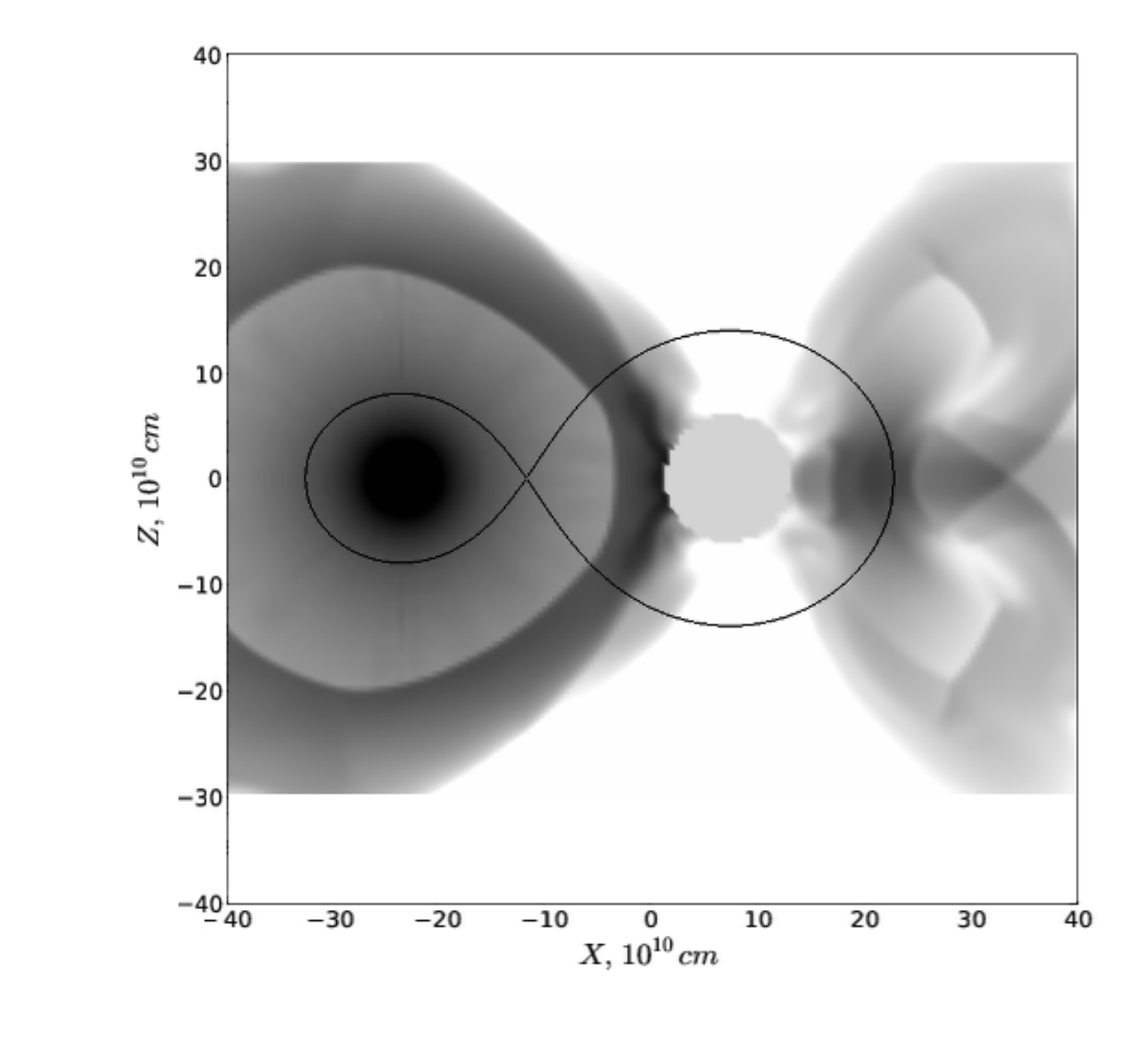}

\caption{Mid plane (X-Y) and perpendicular to it slices through the domains of the simulations performed with GADGET (top panel) and PLUTO (lower panel). The grey scale shows the distribution of the density of the stellar wind matter. Approximate sizes of the companion star and ``accretion volume'' are shown by grey circles. The solid curve shows the Roche lobes of the binary. The wind mass loss rate in these simulations is $\mathrm{10^{-7} M_{\odot}\,yr^{-1}}$.}\label{roche0.4}

\end{figure*}

It is clear that a completely exhaustive model of the accretion flow from a companion star with  a strong stellar wind should include the mechanism of wind generation and its complicated dependence on the strong X-ray illumination by the compact object;  a proper treatment of the accretion disc formation; physics of the X-ray illuminated corona above the accretion disc, etc. Such a comprehensive model would be a complicated and demanding task for an exploratory investigation.
Instead, our approach in this paper is to concentrate  on a simplified model of  mass accretion from the stellar wind. Our aim is to examine whether the variability of the mass accretion rate due to this flow alone can produce characteristic time-scales which may then be observed in the  power density spectra of long period LMXB flux variation. 
	 
For the numerical simulations we use two numerical codes:  a smoothed particle hydrodynamics (SPH) treatment 
of the fluid using the GADGET 2.7 code  \citep{springel_2005}, and the Eulerian grid-based code PLUTO \citep{mignone_2007}. 
These two codes solve the fluid equations via different approaches which have different advantages and disadvantages for our task. Due to the adaptive nature of the SPH method, the computational run time with GADGET is considerably shorter than simulations with PLUTO. On the other hand, the SPH simulations are limited in the their ability to resolve different flow instabilities \citep{agertz_2007} and introduce additional shot noise variations into the accretion rate due to a finite number of particles.

In our calculations we do not simulate the inner structure of the accretion disc which should be formed around the compact object even from the wind accretion.
Simulations of the innermost parts of the accretion flow should include complicated magnetic or turbulent viscosity \citep{balbus_1998} and are beyond the scope of our paper. An internal boundary is placed at an ``accretion radius'' (which we define below), inside of which matter is removed from the simulation volume.

Simulations are conducted in the rotational frame which has its origin at the centre of mass of the system.
The orbital motion of the compact object causes the stellar wind to form a shock wave in front of the compact object and an accretion wake behind it. 
They have an important role in the generation of perturbations in mass accretion rate on the compact object. Therefore, the ``accretion radius'' should not be too big to swallow the region of shock wave formation, otherwise we get a spherically-symmetric accretion.  
We adopt  the ``accretion radius'' to be close to the circularization radius of  captured stellar wind matter, which occurs at $\sim 0.42 R_{\rm 1,L}\sim 6.28\times 10^{10}$ cm $\sim 0.2a$.

\subsection{GADGET}

We have modified the publicly available code GADGET by 
incorporating procedures for adding and deleting particles
from the simulated volume. Particles are added every time step before the domain composition and randomly spread on the star's surface at a  radius slightly less then the Roche lobe radius of the secondary. The rate of appearance of new particles is calculated from the assumed mass loss rate of the star. 

The simulated volume for the GADGET code is a cylinder around the centre of mass with radius $6\times10^{11}$ cm  and semi-height $4.5\times10^{11}$ cm in the direction perpendicular to the orbital plane.
The number of particles is $\sim10^{5}\--6\times10^{5}$.
At the initial state only a small spherical layer around the optical star is filled by particles. The rest of the simulated volume is gradually populated by the stellar wind from the optical star.

The sphere around the compact object is the inner boundary of the computational volume. Particles which pass this boundary contribute to the mass accretion rate of the compact object and are removed from calculations.

\subsection{PLUTO}

 For PLUTO we take a box around the centre of mass with  semi-height $3\times10^{11}$ cm and  semi-width $6\times10^{11}$ cm in both directions along X and Y axis.  A uniform grid with equally sized cells is used. The number of cells is $\mathrm{N_x \times N_y \times N_z}=256\times 256\times128$ (the resolution is $\sim 5\times10^9$ cm).
At the beginning of the simulation the computational volume is filled with low density gas and  stellar wind from the optical star is supplied to the volume. The optical star is simulated as a region with a $1/r^2$ density profile, determined from the wind mass-loss rate and the stellar wind velocity. The stellar wind at this region is purely radial and has a constant velocity, i.e. wind acceleration is not considered.

The accreting region surrounding the compact object is simulated as a region with constant density and constant zero velocity. As this inner boundary condition can potentially have an influence on the parameters of the accreted plasma we have performed simulations with different values for the density of the accreting region:
 equal to the density of undisturbed stellar wind at the distance of compact object and 3.5 times higher. 
The mass accretion rate onto the compact object is calculated as the flow of matter through a sphere surrounding this region.

\section{Results}\label{results}
  
\begin{figure}
\includegraphics[width=\columnwidth]{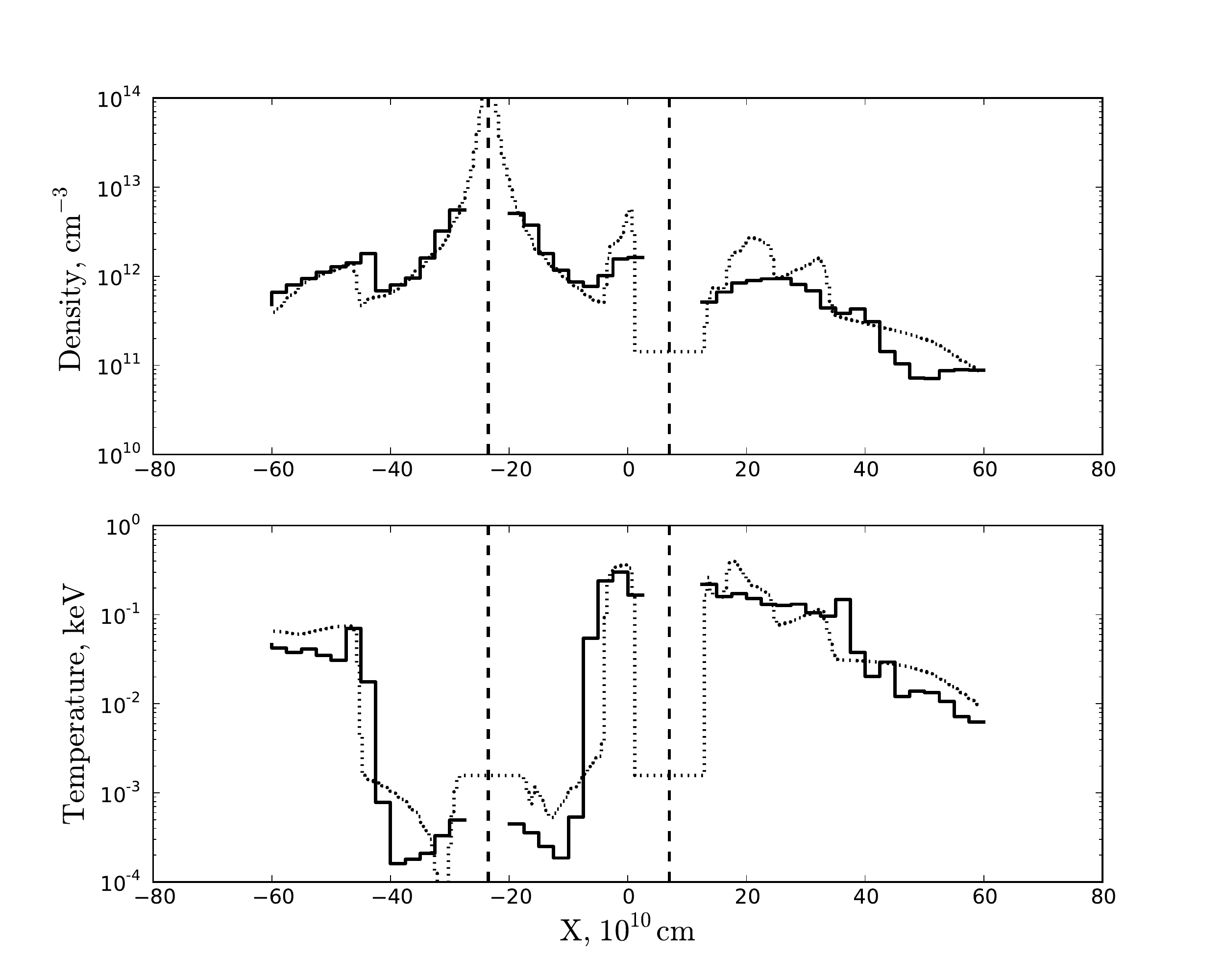}
\caption{Profiles of density and temperature of the wind matter along the line connecting the star's centres for both types of simulations. The results of GADGET simulations are shown by solid histograms, whereas those for the PLUTO simulations are shown by dotted histograms. The mass loss rate is $\mathrm{10^{-7} M_{\odot}\,yr^{-1}}$ in both cases.}\label{profili}

\end{figure}

Snapshots of our simulations performed with the GADGET (top panels) and PLUTO (bottom panels) codes  are presented in Fig. \ref{roche0.4} where slices of density contours in the orbital plane and in the plane perpendicular to it are shown.  
The profiles of densities and temperatures along the line connecting  stars are plotted in Fig. \ref{profili}. Despite of differences between the numerical codes and types of boundary conditions at the position of the compact object, it is clearly seen that the flow patterns are quite similar for both types of codes.

\begin{figure}

\includegraphics[width=\columnwidth]{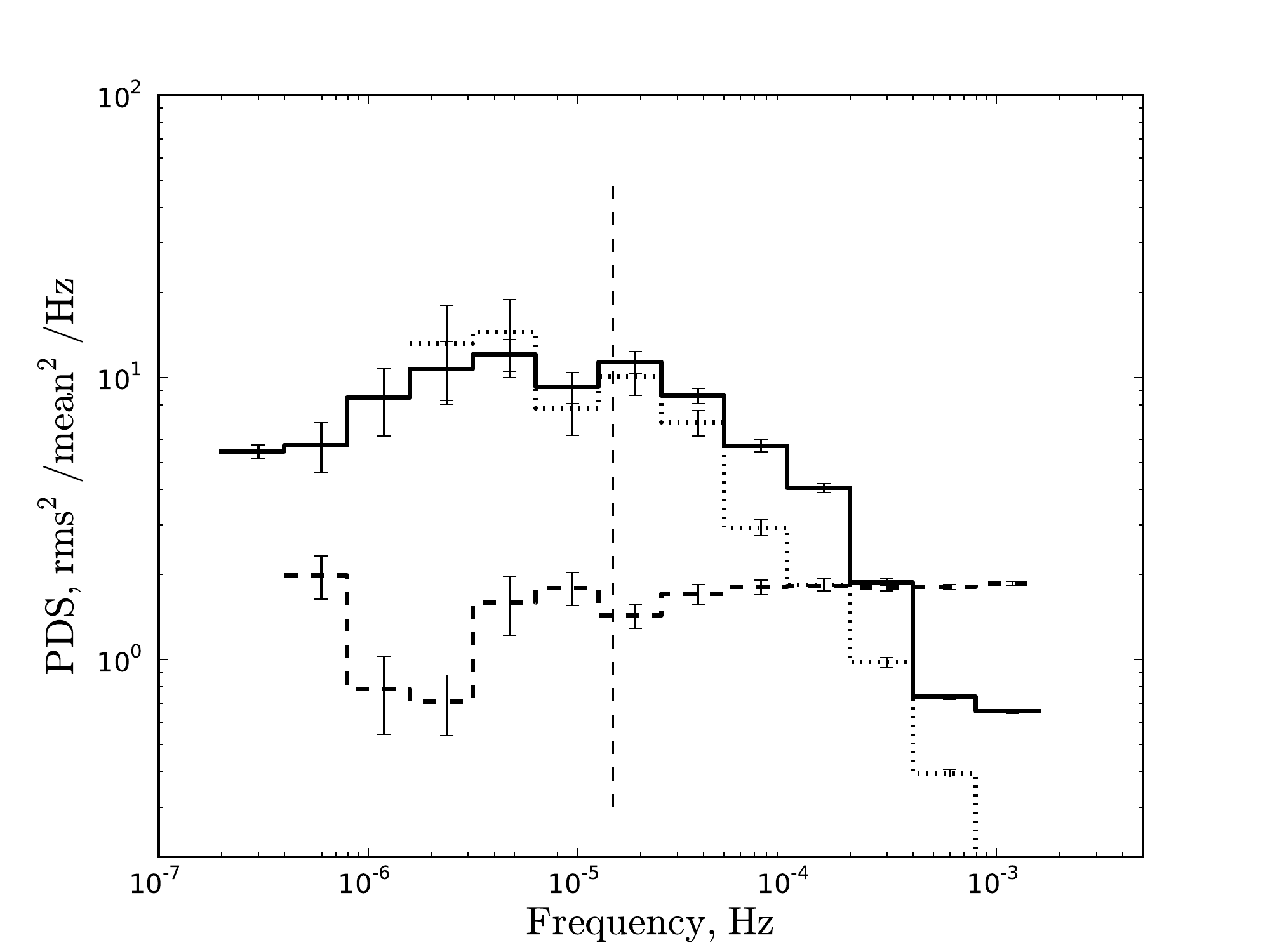}
\caption{Power density spectra for the mass accretion rate variability onto the compact object from the GADGET simulations. The solid histogram shows the PDS  obtained in simulations with $1.2\times 10^5$ particles, whereas the dotted histogram is for $6.6 \times 10^5$ particles. The dashed histogram shows the PDS for a simulation with hydrodynamic interactions turned off. The vertical line shows the orbital frequency of the binary system. The mass loss rate is $\mathrm{5\times 10^{-8} M_{\odot}\,yr^{-1}}$.}\label{pds_gadget}
\end{figure}

The results of our simulations show that the accretion rate calculated as a mass flow rate through the ``accretion radius'' is not constant, but rather demonstrates fluctuations about a mean which appear as band limited noise. The power density spectrum of this variability in the simulations performed with GADGET (using parameters noted in Table 1) is shown in Fig. \ref{pds_gadget} (solid line). The mass loss rate in this simulation  is $\mathrm{5\times 10^{-8} M_{\odot}\,yr^{-1}}$. It is clear that the accretion rate has an excess power at time-scales close to the orbital one (indicated by the vertical dashed line in the figure). Increase of the number of particles in the simulation by a factor of five does not change the shape of the PDS (see Fig. \ref{pds_gadget}, dotted line). The only difference is a decrease in the artificial noise component (shot noise) at high Fourier frequencies.
We have also performed simulations for a mass loss rate of $\mathrm{10^{-7} M_{\odot}\,yr^{-1}}$ and found that it does not alter the general shape of PDS.

Turning off all hydrodynamic forces, which is possible to do in the GADGET simulations (only gravitational and non-inertial forces then remain), we clearly see that the band limited noise in the accretion rate disappears, only the shot noise from the particles remains (see Fig. \ref{pds_gadget}, dashed line). This shows that the mass accretion rate variability is caused by hydrodynamic interactions of matter in the binary.

\begin{figure}

\includegraphics[width=\columnwidth]{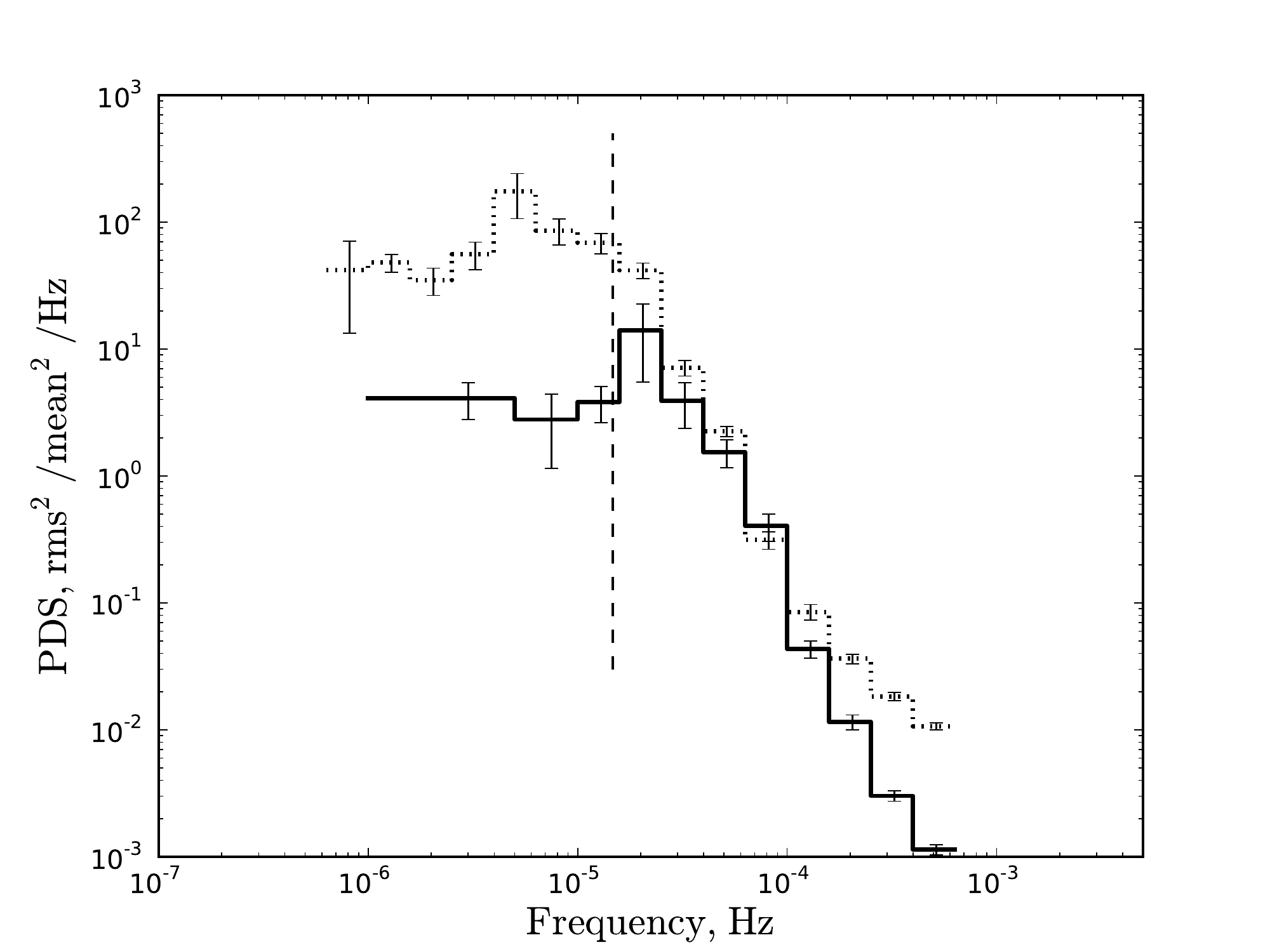}
\caption{Power density spectra for the mass accretion rate variability onto the compact object from the PLUTO simulations. The solid curve shows the PDS obtained from a simulation where the density at the place of the accretion object is equal to the unperturbed density of the stellar wind at the distance of the compact object. The dashed curve shows the PDS obtained in a simulation where the density at the place of the accretion object is 3.5 times higher than in the previous case. The vertical line shows the orbital frequency of the binary system. The mass loss rate is $\mathrm{10^{-7} M_{\odot}\,yr^{-1}}$. }\label{pds_pluto}
\end{figure}

The results of the PLUTO simulation for a mass-loss rate of $\mathrm{10^{-7}M_{\odot}\,yr^{-1}}$ also demonstrate a similar band limited noise in the mass accretion rate (see Fig. \ref{pds_pluto}, the solid line shows the PDS for the case when the density in the accretion region is equal to the unperturbed density of the stellar wind at the distance of the compact object; the dotted line is for the case when the density is 3.5 times higher). The only distinction between the two cases with different densities is in the power at frequencies below the break in the power law, where an increase with increasing density is apparent.

Although the PDS shape obtained in the simulations performed with the different codes is not exactly identical, their overall similarity is clear - both PDSs demonstrate excess of power at frequencies close to the orbital one.

The amplitude of the simulated modulation is a few percent. It is significantly smaller than $\sim10-15$ per cent variability of the observed Sco X-1 X-ray luminosity which we would like to explain. However, we should note that our simulations are exploratory and do not include physical processes which might lead to an increase in the amplitude of fluctuations in the wind mass accretion rate. For example, significant variations of the wind mass loss rate of late type giants (\citealt*{linsky_2000,airapetian_2010} and references therein), cooling and heating of the wind matter which might increase the density contrasts in the wind, and/or interactions with the accretion disc and its corona.  Therefore, our results should be treated as a first step, as an indication of a physical effect and a motivation for the next (more detailed) simulations of the complete accretion flow.

\section{Possible observational appearances of the wind}\label{stellar_wind}

The presence of a relatively  powerful stellar wind in a binary system, which is a necessary condition for our proposed mechanism to work, might lead to additional observational appearances apart from the generation of additional orbital noise in their mass accretion rate (and X-ray luminosities).
For example, stellar winds of single giants and giants in binaries with white dwarfs sometimes can be seen via their bremsstrahlung emission at radio frequencies (see e.g. \citealt{wright_1975,seaquist_1990}), their appearance in optical and infrared emission lines \citep{linsky_2000}, or in infrared dust continuum (\citealt{sopka_1985,ladjal_2010} and references therein). 

On the one hand these methods of stellar wind detection are not suitable for our case because: a) the powerful emission from the compact object dominates at virtually all wavelengths including radio wavelengths due to non-thermal emission of jets \citep{fomalont_2001,mirabel_1999,hannikainen_2009}, and b) illumination of the wind matter should heat it and make it observationally different from the wind of single stars.

\begin{figure}
\includegraphics[width=\columnwidth]{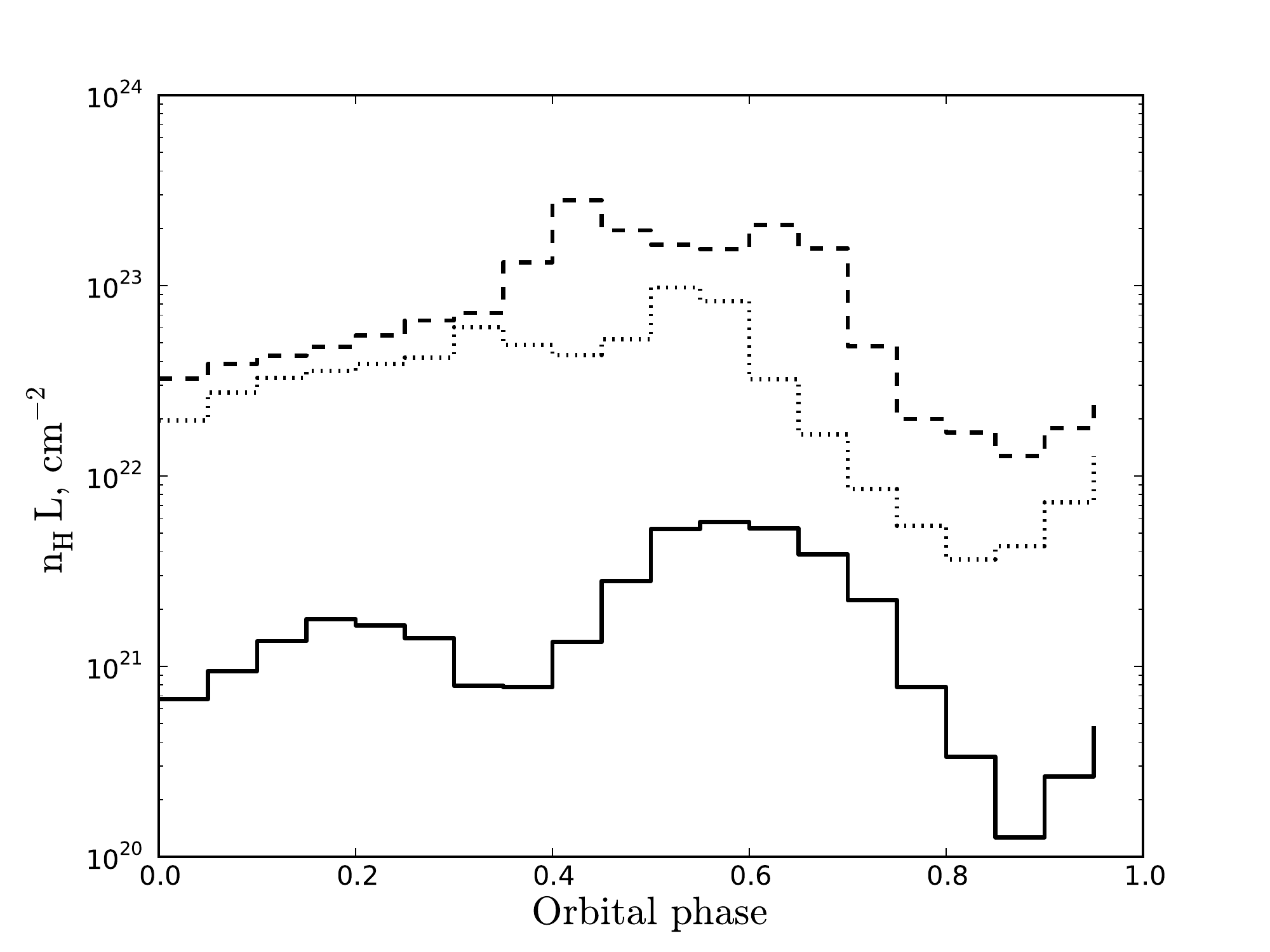}
\caption{Column densities of the wind matter along the line of sight in the binary system for different inclinations, $i=20^\circ,40^\circ,60^\circ$ (solid, dotted and dashed lines correspondingly).}\label{NH_orbital_phase}
\end{figure}

On the other hand the large column density of matter along the line of sight in the binary might lead to absorption of X-rays emerging from the compact object, and the appearance of some emission lines which might be generated in the heated wind matter. 
Our simulations show that with the adopted parameters of the mass loss rate in the wind ($\mathrm{\dot{M}=10^{-7} M_\odot\,yr^{-1}}$), the absorbing column density along the line of sight in the binary can be $n_H L\sim(0.01\--3)\times10^{23}$ cm$^{-2}$  (depending on the binary inclination - see Fig. \ref{NH_orbital_phase}). However, it should be taken into account that the X-ray luminosity of the compact object can significantly diminish the absorbing effect of this matter. The powerful X-ray flux will heat the wind matter to high temperatures, and directly photo-ionize it.

\begin{figure}
\includegraphics[width=\columnwidth]{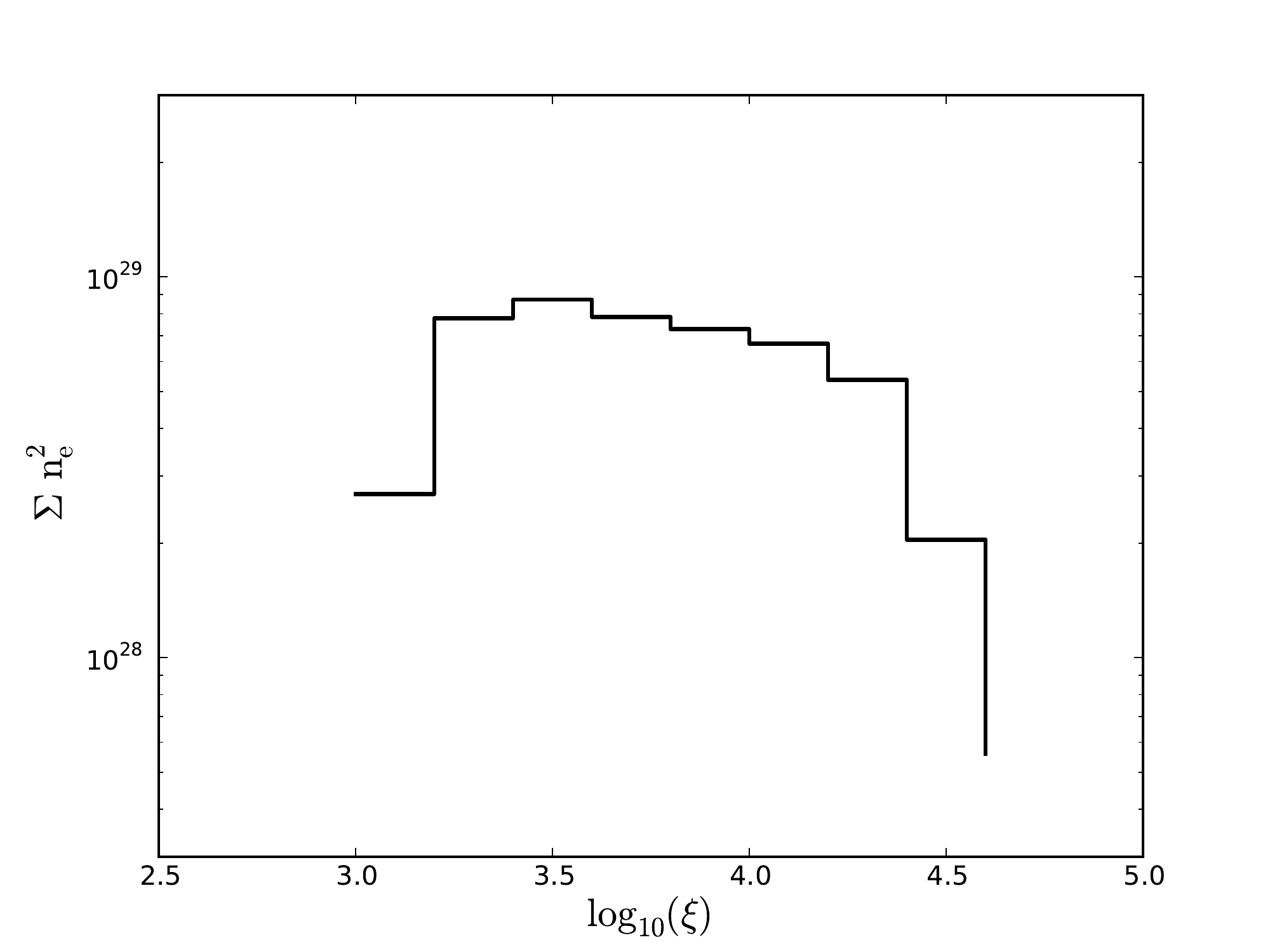}
\caption{Distribution of the ionization parameter in normalized to the volumetric emission measure obtained from the PLUTO simulation.}\label{xi}
\end{figure}

\begin{figure}
\includegraphics[width=\columnwidth]{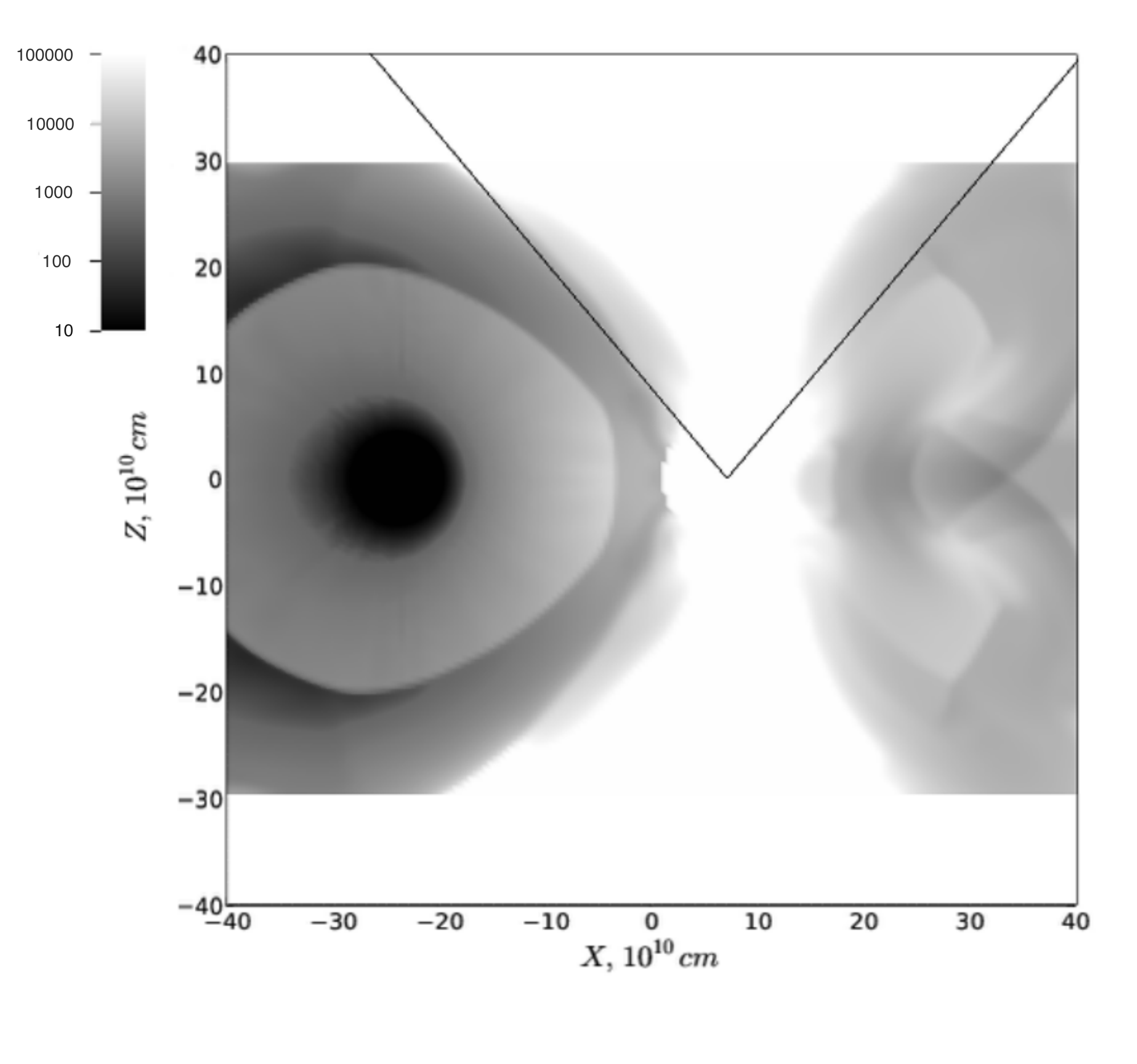}
\caption{Distribution of the ionization parameter obtained from the PLUTO simulation in the XZ plane. The lines show the line of sight corresponding to the inclination angle of the system  $i=40^\circ$.}\label{xi_map}
\end{figure}

In order to check the influence of X-ray emission on absorbing capabilities of the wind we make simple estimates of the spatial distribution of the ionization parameter  $\xi=L_{\rm x}/nR^2$ (where $R$ is the distance of the matter with density $n$ from the X-ray source with luminosity $L_{\rm x}$) in the wind matter over its values using a snapshot from the PLUTO simulations. These calculations are not self-consistent at the moment, i.e. without the influence of heating on the dynamics of the flow. However, they allow us to demonstrate the idea as the ionization parameter distribution has a broad peak around $\log \xi\sim 3.5$ (see Fig. \ref{xi}) which means that almost all elements will be completely ionized and absorbing disabled.  For visualization purpose we also plot the map of ionization parameter in the XZ plane  (Fig. \ref{xi_map}). Lines denote the cone formed by the line of sight during different orbital phases assuming the inclination angle $i=40^\circ$.

\begin{figure}
\includegraphics[width=\columnwidth,bb=39 180 567 700,clip]{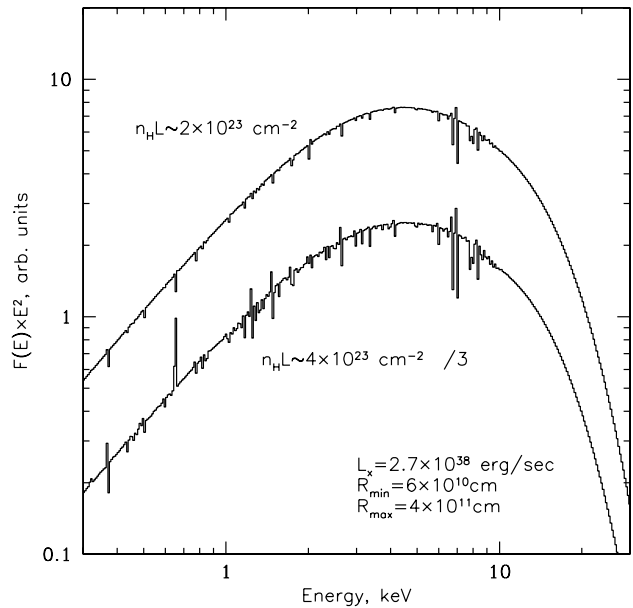}
\caption{Spectra of emission of Sco X-1, seen through a uniform cloud of matter with inner radius $6\times10^{10}$ cm, outer radius $4\times10^{11}$ cm and column density $2\times10^{23}$ cm$^{-2}$ and $4\times10^{23}$ cm$^{-2}$, calculated using the XSTAR code. The spectra are the sum of emission transmitted through the cloud and emission generated by the heated cloud itself.}\label{xstar_sp}
\end{figure}

In order to understand the influence of absorption in photoionized matter of the stellar wind on the X-ray spectrum of Sco X-1 we have made a simple radiative transfer calculation with the help of the XSTAR code \citep{kallman_2001}. We assume that the wind matter forms a spherical cloud with minimal radius $R_{min}=6\times10^{10}$ cm and maximal radius $R_{max}=4\times10^{11}$ cm around the X-ray source which has a luminosity, $L_{\rm x}=2.7 \times 10^{38}\, {\rm erg\,s^{-1}}$ and a spectral shape obtained  by fitting the typical spectrum of Sco X-1. The total absorbing column is taken to be $n_HL=(2\--4)\times 10^{23}$ cm$^{-2}$. Finally, the temperature of the matter in the cloud is calculated iteratively until thermal equilibrium with the radiation is established.
The resulting spectra are shown in Fig. \ref{xstar_sp} (these spectra are the sum of transmitted emission and emission from the cloud itself). It is clear that the influence of the absorbing cloud on the X-ray spectrum is not very strong. The line forest will be completely indistinguishable for X-ray instruments with poor energy resolution but might be detected with high resolution spectroscopy. For example, with the help of the X-ray microcalorimeters on {\it Astro-H} \citep{mitsuda_2012}, especially for the case of very high column density $n_{H}L\sim 4\times10^{23}$ cm$^{-2}$. A more accurate prediction of modifications to the X-ray emission by the stellar wind matter requires a more adequate treatment of radiative cooling and its influence on overall dynamics of the flow, which is beyond the scope of the current paper.
 
From an observational point of view it is worthwhile noting that the variable absorption and emission lines of highly ionized ions in the spectra of Sco X-1 were previously reported by \cite{kahn_1984,christian_1997}. A number of absorption lines of highly ionized ions were detected in the spectrum of the long period LMXBs (presumably with giant or supergiant companion) Cyg X-2 \citep{yao_2009, cabot_2013},  GX 13+1 \citep{ueda_2004}, Cir X-1 \citep{brandt_2000} and GRS 1915+105 \citep{kotani_2000}. Absorption lines in the three latter sources were interpreted as originating from some kind of accretion disc wind \citep[see e.g.][]{begelman_1983}. Our results show that the presence of off-orbital plane matter from the stellar wind of the giant companion can also contribute to a set of absorption and emission lines in the spectra of long period LMXBs. 

\section{Conclusions}\label{conclusions}

Three dimensional hydrodynamic simulations have been used to investigate the time variability of the wind accretion rate in an LMXB system with a late type giant companion. 
Our results may by summarized as follows.

\begin{itemize}

\item We show that the accretion rate from the stellar wind of a late type giant in an LMXB is intrinsically time variable  with a characteristic time-scale close to the orbital period of the system even in the case of zero eccentricity orbit. 

\item We demonstrate that this variability is caused by a hydrodynamical interaction of the stellar wind via the undeveloped turbulent motion (perturbed motion without significant vorticities) of the matter around compact object.

\item Our simulations show  variations of only a few percent in the mass accretion rate from the stellar wind. This is lower than the observed 10-15 per cent level of variability. However, we anticipate that this is caused due to the simplicity of our model and the inclusion of additional physics in the simulation might lead to an increase in the variability amplitude.

\item  In this work we use a companion star mass-loss rate of $(0.5-1)\times10^{-7}~M_\odot$/year. This value is significantly higher than the typical mass loss rate of a single giant star of this type. However, examples of even higher mass loss rates for giant stars in binaries do exist. Although the mechanism of wind mass loss enhancement is not clear, it might be related to the strong illumination of the secondary by X-ray emission of the compact object.

\item We estimate some potential observational signatures of the stellar wind in the binary system, finding that similar emission and absorption line features to an accretion disc corona/wind may be produced.
\end{itemize}

In summary, the proposed mechanism for the generation of long time-scale variability (i.e. on the order of the orbital period) in the observed X-ray luminosity of persistent LMXBs with giant companions appears viable and should be explored further with more detailed numerical and observational studies.

\section*{Acknowledgements}

EF and MR acknowledge support by grant NSh-5603.2012.2, grant RFBR 13-02-00741, program P19 of Presidium of the Russian Academy of Sciences (RAS) and program OFN17 of the Division of Physical Sciences of the RAS.  EF also thanks Go8 Fellowship and  grant 20FI20\_135263 of National Swiss Fond.


\begin{thebibliography}{}

\bibitem[\protect\citeauthoryear{Agertz et~al.,}{Agertz
  et~al.}{2007}]{agertz_2007}
Agertz O.  et~al., 2007, ApJ, 380, 963

\bibitem[\protect\citeauthoryear{Airapetian, Carpenter \& Ofman}{Airapetian
  et~al.}{2010}]{airapetian_2010}
Airapetian V.,  Carpenter K.,    Ofman L.,  2010, ApJ, 723, 1210

\bibitem[\protect\citeauthoryear{Arons}{Arons}{1973}]{arons_1973}
Arons J.,  1973, ApJ, 184, 539

\bibitem[\protect\citeauthoryear{Balbus \& Hawley}{Balbus \&
  Hawley}{1998}]{balbus_1998}
Balbus S.,  Hawley J.,  1998, Rev. Mod. Phys., 70, 1

\bibitem[\protect\citeauthoryear{Bandyopadhyay, Shahbaz, Charles \&
  Naylor}{Bandyopadhyay et~al.}{1999}]{bandy_1999}
Bandyopadhyay R.,  Shahbaz T.,  Charles P.,    Naylor T.,  1999, MNRAS, 306,
  417

\bibitem[\protect\citeauthoryear{Bandyopadhyay, Shahbaz, Charles, van Kerkwijk
  \& Naylor}{Bandyopadhyay et~al.}{1997}]{bandy_1997}
Bandyopadhyay R.,  Shahbaz T.,  Charles P.~A.,  van Kerkwijk M.~H.,    Naylor
  T.,  1997, MNRAS, 285, 718

\bibitem[\protect\citeauthoryear{Basko \& Sunyaev}{Basko \&
  Sunyaev}{1973}]{basko_1973}
Basko M.,  Sunyaev R.,  1973, Ap\&SS, 23, 117

\bibitem[\protect\citeauthoryear{Bayless et~al.}{Bayless et~al.}{2010}]{bayless_2010}
Bayless A.,  Robinson E.,  Hynes R.,  Ashcraft T.,    Cornell M.,  2010, A\&A,
  709, 251

\bibitem[\protect\citeauthoryear{Begelman, McKee \& Shields}{Begelman
  et~al.}{1983}]{begelman_1983}
Begelman M.,  McKee C.,    Shields G.,  1983, ApJ, 271, 70

\bibitem[\protect\citeauthoryear{Biermann}{Biermann}{1971}]{biermann_1971}
Biermann P.,  1971, A\&A, 10, 205

\bibitem[\protect\citeauthoryear{Bogdanov \& Taranova}{Bogdanov \&
  Taranova}{2008}]{bogdanov_2008}
Bogdanov M.,  Taranova O.,  2008, Astron. Rep., 52, 403

\bibitem[\protect\citeauthoryear{Boirin et~al.}{Boirin et~al.}{2005}]{boirin_2005}
Boirin L.,  M\'endez M.,  D\'iaz~Trigo M.,  Parmar A.,    Kaastra J.,  2005,
  A\&A, 436, 195

\bibitem[\protect\citeauthoryear{Boirin et~al.}{Boirin
  et~al.}{2004}]{boirin_2004}
Boirin L.,  Parmar A.,  Barret D.,    Paltani S.,  2004, Nuclear Physics B
  (Proc. Suppl.), 132, 506

\bibitem[\protect\citeauthoryear{Brandt \& Schulz}{Brandt \&
  Schulz}{2000}]{brandt_2000}
Brandt W.,  Schulz N.,  2000, ApJ, 544, 123

\bibitem[\protect\citeauthoryear{Cabot, Wang \& Yao}{Cabot
  et~al.}{2013}]{cabot_2013}
Cabot S.,  Wang D.,    Yao Y.,  2013, MNRAS, 431, 511

\bibitem[\protect\citeauthoryear{Chakrabarty, van Kerkwijk \&
  Larkin}{Chakrabarty et~al.}{1998}]{chakrabarty_1998}
Chakrabarty D.,  van Kerkwijk M.,    Larkin J.,  1998, ApJ, 497, 39

\bibitem[\protect\citeauthoryear{Christian \& Swank}{Christian \&
  Swank}{1997}]{christian_1997}
Christian D.,  Swank J.,  1997, ApJS, 109, 177

\bibitem[\protect\citeauthoryear{Churazov, Gilfanov \& Revnivtsev}{Churazov
  et~al.}{2001}]{churazov_2001}
Churazov E.,  Gilfanov M.,    Revnivtsev M.,  2001, MNRAS, 321, 759

\bibitem[\protect\citeauthoryear{Cottam et~al.}{Cottam
  et~al.}{2001}]{cottam_2001}
Cottam J.,  Sako M.,  Kahn S.,  Paerels F.,    Liedahl D.,  2001, A\&A, 365,
  277

\bibitem[\protect\citeauthoryear{{D'Ai}, Zycki, Di~Salvo, Iaria, Lavagetto \&
  Robba}{{D'Ai} et~al.}{2007}]{dai_2007}
{D'Ai} A.,  Zycki P.,  Di~Salvo T.,  Iaria R.,  Lavagetto G.,    Robba N.,
  2007, ApJ, 667, 441

\bibitem[\protect\citeauthoryear{de Val-Borro, Karovska \&
  Sasselov}{de~Val-Borro et~al.}{2009}]{valborro_2009}
de Val-Borro M.,  Karovska M.,    Sasselov D.,  2009, ApJ, 700, 1148

\bibitem[\protect\citeauthoryear{Dermine, Jorissen, Siess \&
  Frankowski}{Dermine et~al.}{2009}]{dermine_2009}
Dermine T.,  Jorissen A.,  Siess L.,    Frankowski A.,  2009, A\&A, 507, 891

\bibitem[\protect\citeauthoryear{Faraggiana \& Hack}{Faraggiana \&
  Hack}{1969}]{faraggiana_1969}
Faraggiana R.,  Hack M.,  1969, Ap\&SS, 3, 205

\bibitem[\protect\citeauthoryear{Filippova, Revnivtsev \& Lutovinov}{Filippova
  et~al.}{2008}]{filippova_2008}
Filippova E.,  Revnivtsev M.,    Lutovinov A.,  2008, Ast. Lett., 34, 797

\bibitem[\protect\citeauthoryear{Fomalont, Geldzahler \& Bradshaw}{Fomalont
  et~al.}{2001}]{fomalont_2001}
Fomalont E.,  Geldzahler B.,    Bradshaw C.,  2001, ApJ, 558, 283

\bibitem[\protect\citeauthoryear{Frank, King \& Raine}{Frank
  et~al.}{2002}]{frank_2002}
Frank J.,  King A.,    Raine D.,  2002, Accretion power in astrophysic.
Cambridge university press

\bibitem[\protect\citeauthoryear{Gilfanov \& Arefiev}{Gilfanov \&
  Arefiev}{2005}]{gilfanov_2005}
Gilfanov M.,  Arefiev V.,  2005, preprint (astro-ph/0501215)

\bibitem[\protect\citeauthoryear{Hannikainen et~al.}{Hannikainen
  et~al.}{2009}]{hannikainen_2009}
Hannikainen D. et~al.,  2009, MNRAS, 397, 569

\bibitem[\protect\citeauthoryear{Iben, Tutukov \& Fedorova}{Iben
  et~al.}{1997}]{iben_1997}
Iben I.,  Tutukov A.,    Fedorova A.,  1997, ApJ, 486, 955

\bibitem[\protect\citeauthoryear{Ingram \& van der Klis}{Ingram \& van der Klis}{2013}]{ingram_2013}
Ingram A., van der Klis M.,   2013, MNRAS, 434, 1476

\bibitem[\protect\citeauthoryear{Jimenez-Garate, Schulz \&
  Marshall}{Jimenez-Garate et~al.}{2003}]{jimenez_2003}
Jimenez-Garate M.,  Schulz N.,    Marshall H.,  2003, ApJ, 590, 432

\bibitem[\protect\citeauthoryear{Kahn, Seward \& Chlebowski}{Kahn
  et~al.}{1984}]{kahn_1984}
Kahn S.,  Seward F.,    Chlebowski T.,  1984, ApJ, 283, 286

\bibitem[\protect\citeauthoryear{Kallman et~al.}{Kallman
  et~al.}{2003}]{kallman_2003}
Kallman T.,  Angelini L.,  Boroson B.,    Cottam J.,  2003, ApJ, 583, 861

\bibitem[\protect\citeauthoryear{Kallman \& Bautista}{Kallman \&
  Bautista}{2001}]{kallman_2001}
Kallman T.,  Bautista M.,  2001, ApJS, 133, 221

\bibitem[\protect\citeauthoryear{Kotani, Ebisawa, Dotani, Inoue, Nagase, Tanaka
  \& Ueda}{Kotani et~al.}{2000}]{kotani_2000}
Kotani T.,  Ebisawa K.,  Dotani T.,  Inoue H.,  Nagase F.,  Tanaka Y.,    Ueda
  Y.,  2000, ApJ, 539, 413

\bibitem[\protect\citeauthoryear{Kotov, Churazov \& Gilfanov}{Kotov
  et~al.}{2001}]{kotov_2001}
Kotov O.,  Churazov E.,    Gilfanov M.,  2001, MNRAS, 327, 799

\bibitem[\protect\citeauthoryear{Ladjal, Justtanont, Groenewegen, Blommaert,
  Waelkens \& Barlow}{Ladjal et~al.}{2010}]{ladjal_2010}
Ladjal D.,  Justtanont K.,  Groenewegen M.,  Blommaert J.,  Waelkens C.,
  Barlow M.,  2010, A\&A, 513, 53

\bibitem[\protect\citeauthoryear{Linsky et~al.}{Linsky et~al.}{2000}]{linsky_2000}
Linsky J.,  Harper G.,  Valenti J.,  Bennett P.,    Brown A.,  2000, IAUS, 117,
  303

\bibitem[\protect\citeauthoryear{Lyubarskii}{Lyubarskii}{1997}]{lyubarskii_1997}
Lyubarskii Y.,  1997, MNRAS, 292, 679

\bibitem[\protect\citeauthoryear{Matsuda, Inoue \& Sawada}{Matsuda
  et~al.}{1987}]{matsuda_1987}
Matsuda T.,  Inoue M.,    Sawada K.,  1987, MNRAS, 226, 785

\bibitem[\protect\citeauthoryear{Mignone, Bodo, Massaglia, Matsakos, Tesileanu,
  Zanni \& Ferrari}{Mignone et~al.}{2007}]{mignone_2007}
Mignone A.,  Bodo G.,  Massaglia S.,  Matsakos T.,  Tesileanu O.,  Zanni C.,
  Ferrari A.,  2007, ApJS, 170, 228

\bibitem[\protect\citeauthoryear{Mirabel \& Rodriguez}{Mirabel \&
  Rodriguez}{1999}]{mirabel_1999}
Mirabel I.,  Rodriguez L.,  1999, ARA\&A, 37, 409

\bibitem[\protect\citeauthoryear{Mitsuda, Kelley, Boyce, Brown, Costantini,
  DiPirro, Ezoe \& Fujimoto}{Mitsuda et~al.}{2012}]{mitsuda_2012}
Mitsuda K.,  Kelley R.,  Boyce K.,  Brown G.,  Costantini E.,  DiPirro M.,
  Ezoe Y.,    Fujimoto R.,  2012, JLTP, 167, 795

\bibitem[\protect\citeauthoryear{Nagae et~al.}{Nagae et~al.}{2004}]{nagae_2004}
Nagae T.,  Oka K.,  Matsuda T.,  Fujiwara H.,  Hachisu I.,    Boffin H.,  2004,
  A\&A, 419, 335

\bibitem[\protect\citeauthoryear{Reimers}{Reimers}{1975}]{reimers_1975}
Reimers D.,  1975, MSRSL, 7, 369

\bibitem[\protect\citeauthoryear{Reimers}{Reimers}{1977}]{reimers_1977}
Reimers D.,  1977, A\&A, 61, 217

\bibitem[\protect\citeauthoryear{Schroder \& Cuntz}{Schroder \&
  Cuntz}{2005}]{schroder_2005}
Schroder K.,  Cuntz M.,  2005, ApJ, 630, 73

\bibitem[\protect\citeauthoryear{Seaquist \& Taylor}{Seaquist \&
  Taylor}{1990}]{seaquist_1990}
Seaquist E.,  Taylor A.,  1990, ApJ, 349, 313

\bibitem[\protect\citeauthoryear{Shakura \& Sunyaev}{Shakura \&
  Sunyaev}{1973}]{shakura_1973}
Shakura N.,  Sunyaev R.,  1973, A\&A, 24, 337

\bibitem[\protect\citeauthoryear{Sibgatullin \& Sunyaev}{Sibgatullin \&
  Sunyaev}{2000}]{sibgatullin_2000}
Sibgatullin N.,  Sunyaev R.,  2000, Astron. Lett, 26, 699

\bibitem[\protect\citeauthoryear{Skopal, Bode, Crocker, Drechsel, Eyres \&
  Komzik}{Skopal et~al.}{2002}]{skopal_2002}
Skopal A.,  Bode M.,  Crocker M.,  Drechsel H.,  Eyres S.,    Komzik R.,  2002,
  MNRAS, 335, 1109

\bibitem[\protect\citeauthoryear{Sopka, Hildebrand, Jaffe, Gatley, Roellig,
  Werner, Jura \& Zuckerman}{Sopka et~al.}{1985}]{sopka_1985}
Sopka R.,  Hildebrand R.,  Jaffe D.,  Gatley I.,  Roellig T.,  Werner M.,  Jura
  M.,    Zuckerman B.,  1985, ApJ, 294, 242

\bibitem[\protect\citeauthoryear{Springel}{Springel}{2005}]{springel_2005}
Springel V.,  2005, MNRAS, 364, 1105

\bibitem[\protect\citeauthoryear{Steeghs \& Casares}{Steeghs \&
  Casares}{2002}]{steeghs_2002}
Steeghs D.,  Casares J.,  2002, ApJ, 568, 273

\bibitem[\protect\citeauthoryear{Theuns, Boffin \& Jorissen}{Theuns
  et~al.}{1996}]{theuns_1996}
Theuns T.,  Boffin H.,    Jorissen A.,  1996, MNRAS, 280, 1264

\bibitem[\protect\citeauthoryear{Ueda, Murakami, Yamaoka, Dotani \&
  Ebisawa}{Ueda et~al.}{2004}]{ueda_2004}
Ueda Y.,  Murakami H.,  Yamaoka K.,  Dotani T.,    Ebisawa K.,  2004, ApJ, 609,
  325

\bibitem[\protect\citeauthoryear{Uttley \& McHardy}{Uttley \&
  McHardy}{2001}]{uttley_2001}
Uttley P.,  McHardy I.,  2001, MNRAS, 323, 26

\bibitem[\protect\citeauthoryear{Wright \& Barlow}{Wright \&
  Barlow}{1975}]{wright_1975}
Wright A.,  Barlow M.,  1975, MNRAS, 170, 41

\bibitem[\protect\citeauthoryear{Yao, Schulz, Gu, Nowak \& Canizares}{Yao
  et~al.}{2009}]{yao_2009}
Yao Y.,  Schulz N.,  Gu M.,  Nowak M.,    Canizares R.,  2009, ApJ, 696, 1418

\end{thebibliography}
\end{document}